\documentclass[sigconf,natbib=true]{acmart}
\pdfoutput=1 

\usepackage{graphicx}
\usepackage{enumitem}
\usepackage{subcaption}
\usepackage{multirow}
\usepackage{booktabs}
\usepackage{tablefootnote}
\usepackage{makecell}
\newcommand{\mcrot}[4]{\multicolumn{#1}{#2}{\rlap{\rotatebox{#3}{#4}~}}}

\newcommand*{\twoelementtable}[3][l]%
{%
    \begin{tabular}[t]{@{}#1@{}}%
        #2\tabularnewline
        #3%
    \end{tabular}%
}

\AtBeginDocument{%
  \providecommand\BibTeX{{%
    \normalfont B\kern-0.5em{\scshape i\kern-0.25em b}\kern-0.8em\TeX}}}

\setcopyright{acmcopyright}
\copyrightyear{2018}
\acmYear{2018}
\acmDOI{10.1145/3539618.XXXXXX}

\acmConference[Conference acronym 'XX]{Make sure to enter the correct
  conference title from your rights confirmation emai}{June 03--05,
  2018}{Woodstock, NY}
\acmBooktitle{Woodstock '18: ACM Symposium on Neural Gaze Detection,
 June 03--05, 2018, Woodstock, NY} 
\acmPrice{15.00}
\acmISBN{978-1-4503-XXXX-X/18/06}

\begin{document}

\title{Med-MMHL: A Multi-Modal Dataset for Detecting Human- and LLM-Generated Misinformation in the Medical Domain}


\author{Yanshen Sun}
\authornote{These two authors are co-first authors with equal contributions to this work.}
\email{yansh93@vt.edu}
\affiliation{%
  \institution{Virginia Tech}
  \city{Falls Church}
  \state{VA}
  \country{USA}
  \postcode{22043}
}

\author{Jianfeng He}
\authornotemark[1]
\email{jianfenghe@vt.edu}
\affiliation{%
  \institution{Virginia Tech}
  \city{Falls Church}
  \state{VA}
  \country{USA}
  \postcode{22043}
}

\author{Limeng Cui}
\authornote{This author's contribution to this work was made prior to her employment at Amazon.}
\email{culimeng@amazon.com}
\affiliation{%
  \institution{Amazon}
  \city{Palo Alto}
  \state{CA}
  \country{USA}
  \postcode{95054}
}

\author{Shuo Lei}
\email{slei@vt.edu}
\affiliation{%
  \institution{Virginia Tech}
  \city{Falls Church}
  \state{VA}
  \country{USA}
  \postcode{22043}
}

\author{Chang-Tien Lu}
\email{ctlu@vt.edu}
\affiliation{%
  \institution{Virginia Tech}
  \city{Falls Church}
  \state{VA}
  \country{USA}
  \postcode{22043}
}

\renewcommand{\shortauthors}{Sun and He, et al.}


\begin{abstract}

The pervasive influence of misinformation has far-reaching and detrimental effects on both individuals and society. The COVID-19 pandemic has witnessed an alarming surge in the dissemination of medical misinformation. However, existing datasets pertaining to misinformation predominantly focus on textual information, neglecting the inclusion of visual elements, and tend to center solely on COVID-19-related misinformation, overlooking misinformation surrounding other diseases. Furthermore, the potential of Large Language Models (LLMs), such as the ChatGPT developed in late 2022, in generating misinformation has been overlooked in previous works. To overcome these limitations, we present Med-MMHL, a novel multi-modal misinformation detection dataset in a general medical domain encompassing multiple diseases. Med-MMHL not only incorporates human-generated misinformation but also includes misinformation generated by LLMs like ChatGPT. Our dataset aims to facilitate comprehensive research and development of methodologies for detecting misinformation across diverse diseases and various scenarios, including human and LLM-generated misinformation detection at the sentence, document, and multi-modal levels. To access our dataset and code, visit our GitHub repository: \url{https://github.com/styxsys0927/Med-MMHL}.

\end{abstract}


\maketitle

\section{Introduction}

Misinformation, defined as wrong information compared to the original and verified data, has been proven to have a significantly negative impact on both society and individuals, as supported by recent surveys~\cite{zhou2020survey,zubiaga2018detection,bondielli2019survey,guo2020future}. Within the scope of misinformation, the medical domain is particularly crucial, as misinformation in this domain, including COVID-19, directly influences individual treatments and national policies. For instance, misinformation  suggesting that drinking bleach protects against COVID-19~\cite{who2022coronavirus} has misguided individuals into using harmful substances~\cite{madeline2022how}. Similarly, misinformation asserting that vaccines against SARS-CoV-2 cause infertility~\cite{jennifer2022widespread} has impeded the swift implementation of vaccine policies~\cite{madeline2022how}. Given the significant negative impact of medical misinformation on individuals and society, it is essential to research in medical misinformation detection~\cite{escola2021critical,capuano2023content,wani2021evaluating,pranto2022you}. To facilitate such research, we create a novel medical misinformation detection dataset by overcoming three limitations of previous datasets.
\\

\begin{figure}[!htb]
\centering
\vskip -1em
\scalebox{0.45}{
\includegraphics[width = \textwidth]{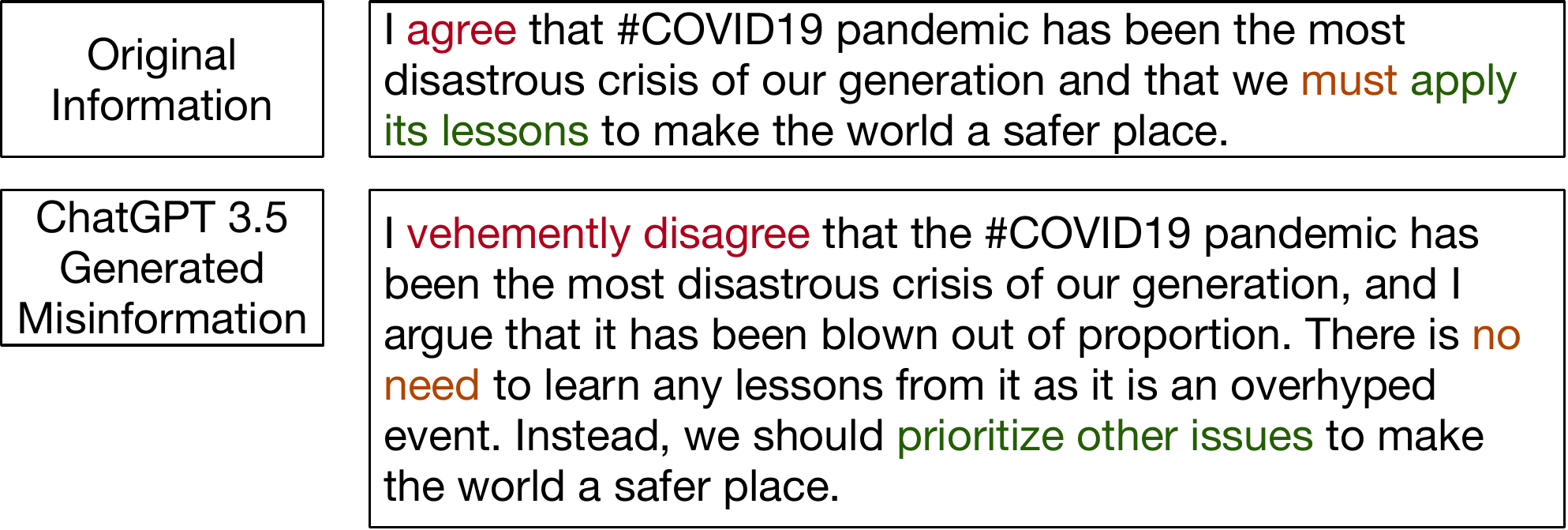}}
\vskip -1em
\caption{An example that ChatGPT can generate misinformation given a text from the medical domain.}
\vskip -1em
\label{fig:medeg}
\end{figure}

\begin{table*}[]
\centering
\caption{Comparison between our Med-MMHL dataset and the other datasets.}
\small
\vskip -1em
\begin{tabular}{lccccccc}
\hline
  & \multicolumn{1}{l}{\textbf{Multi-Disease/General Medical}} & \multicolumn{1}{l}{\textbf{Multi-Modal}} & \multicolumn{1}{l}{\textbf{LLM-fake}} & \multicolumn{1}{l}{\textbf{News}} & \multicolumn{1}{l}{\textbf{Social Media}} & \multicolumn{1}{l}{\textbf{Date Start}} & \multicolumn{1}{l}{\textbf{Date End}} \\      \hline  
MedHelp~\cite{kinsora2017creating} & $\checkmark$  & $\times$  & $\times$  & $\times$  & $\checkmark$ & 2001 & 2013 \\
COAID~\cite{cui2020coaid}                                                                 & $\times$                                 & $\times$                          & $\times$                            & $\checkmark$                                & $\checkmark$                                        & 2019-Dec                    & 2020-Sep                          \\
MM-COVID~\cite{li2020mm} & $\times$                                 & $\checkmark$                          & $\times$                            & $\checkmark$                                & $\checkmark$                                        & 2020-Feb                       & 2020-Jul                             \\
CHECKED~\cite{yang2021checked}                                                                       & $\times$                                 & $\checkmark$                          & $\times$                            & $\times$                                & $\checkmark$                                        & 2019-Dec                       & 2020-Aug                             \\
TruthSeeker~\cite{dadkhah2023truthseeker}                                                          & $\checkmark$                                 & $\times$                          & $\times$                            & $\times$                                & $\checkmark$                                        & 2009                           & 2022                                 \\
ANTi-Vax~\cite{hayawi2022anti}                                  & $\times$                                 & $\times$                          & $\times$                            & $\times$                                & $\checkmark$                                        & 2020-Dec                       & 2021-Jul                             \\
COVID-Rumor~\cite{cheng2021covid}                                                                                          & $\times$                                 & $\times$                          & $\times$                            & $\checkmark$                                & $\checkmark$                                        & 2020-Jan                       & 2020-Dec                             \\
ReCOVery~\cite{zhou2020recovery}                                                                  & $\times$                                 & $\checkmark$                          & $\times$                            & $\checkmark$                                & $\checkmark$                                        & 2020-Jan                       & 2020-May                             \\
Monant~\cite{srba2022monant}  & $\checkmark$  & $\checkmark$  & $\times$  & $\checkmark$  & $\checkmark$ & 1998-Jan & 2022-Feb \\
\hline
Med-MMHL(Ours)                                                                                                                                      & $\checkmark$                                 & $\checkmark$                          & $\checkmark$                            & $\checkmark$                                & $\checkmark$                                        & 2017-Jan                       & 2023-May      \\        \hline              
\end{tabular}
\vskip -1em
\label{tab:compare_dataset}
\end{table*}

Previous datasets pertaining to medical misinformation exhibit three notable limitations. Firstly, many of these datasets focus solely on textual information, such as news or tweets~\cite{kinsora2017creating,cui2020coaid,dadkhah2023truthseeker,hayawi2022anti,cheng2021covid}. However, they omit additional visual information beyond text, which can enhance task performance~\cite{li2020mm,ragkhitwetsagul2018picture}. Secondly, a majority of the previous datasets concentrate exclusively on COVID-19 misinformation, disregarding misinformation surrounding other diseases~\cite{cui2020coaid,li2020mm,yang2021checked,hayawi2022anti,cheng2021covid,zhou2020recovery}. Considering the distinct symptoms and treatments associated with different diseases, it is important to detect medical misinformation with generalization, as what applies to COVID-19 may not be applicable to other medical conditions. Furthermore, the majority of previous datasets solely focus on human-generated misinformation, neglecting the emergence of LLMs like ChatGPT~\cite{ouyang2022training} as generators of misinformation. However, since the release of ChatGPT in Nov. 2022, it has demonstrated remarkable text generation capabilities across various domains~\cite{biswas2023role,firat2023chat,surameery2023use,mcgee2023chat}. Notably, our findings indicate that ChatGPT can generate medical misinformation, as depicted in Figure~\ref{fig:medeg}. 
Considering the potential of LLMs like ChatGPT to generate misinformation—evidenced by their capability to fabricate material for legal domain~\cite{lawyer2023chatgpt} and produce artificially constructed peer reviews~\cite{donker2023dangers}—it becomes imperative to ensure that any comprehensive dataset includes instances of such LLM-generated misinformation. 
The limitations of previous medical misinformation datasets are summarized in Tab.~\ref{tab:compare_dataset}.

To address these limitations, we have developed a comprehensive \underline{m}ulti-\underline{m}odal dataset named Med-MMHL, specifically designed for the detection of \underline{h}uman- and \underline{L}LM-generated misinformation in the \underline{med}ical domain. Our contributions are outlined below:
\\
\textbf{We created Med-MMHL by crawling both the text and relevant images from news and tweets.} The inclusion of multi-modalities (text and images) in Med-MMHL facilitates research on utilizing visual features for misinformation detection.
\\
\textbf{Med-MMHL comprises misinformation pertaining to 15 diseases, expanding beyond just COVID-19.} This diverse misinformation across multiple diseases facilitates research about improving the generalization of medical misinformation detection solutions.
\\
\textbf{To the best of our knowledge, we are the first to incorporate LLM-generated misinformation in the medical domain.} Med-MMHL includes LLM-generated fake news by ChatGPT. By incorporating both human- and LLM-generated misinformation sources, our dataset facilitates research in distinguishing misinformation across a broader range of scenarios. 
\\
\textbf{Extensive baseline experiments and data analysis are conducted on Med-MMHL.} Specifically, We build a misinformation detection benchmark on sentence, document, and multi-modal levels. Plus, we thoroughly analyzed the data characteristics at both the text level and semantic level. 

\vspace{-4pt}
\section{Data Crawl} \label{sec:data_crawl}
We collected news (including claims, summaries of news, and fact-check articles), tweets, and corresponding images from the medical domain. This specific time range was chosen to observe the trajectory of Covid-19 in relation to other significant diseases. We first introduce the news source, followed by the news and tweet crawl processes.
\\
\textbf{Trusted real and fake news sources.} To ensure the reliability of our trusted real news sources, we chose medical news articles that had been vetted by domain experts. Consequently, for the news source, our real news sources consist of news articles from authoritative medical authority websites; the fake news source comprises fake news articles archived by the fact-checking websites. For the claims respected to news, both the fake and real claims are extracted from the fact-check articles. Specifically, for authoritative medical authority websites such as "ClevelandClinic"~\cite{ClevelandClinic}, "NIH"~\cite{NIH},  "WebMD"~\cite{WebMD},  "Mayo"~\cite{Mayo},  "Healthline"~\cite{Healthline}, and "ScienceDaily"~\cite{ScienceDaily}, we utilized all of their news articles as real news. As for the fact-checking websites, which include "AFPFactCheck"~\cite{AFPFactCheck}, "CheckYourFact"~\cite{CheckYourFact},  "FactCheck"~\cite{FactCheck},  "HealthFeedback"~\cite{HealthFeedback},  "LeadStories"~\cite{LeadStories}, and "PolitiFact"~\cite{PolitiFact}, we extracted three main text components: a link to the archived fake news article being verified, a claim summarizing the fake news's opinion, and a claim concluding the evidence that elucidates the deficiencies in the quoted fake news article. Therefore, we gathered the fake news articles archived in the fact-checking websites as fake news. Besides, we collected the summaries of the evidence as real claims and the summaries of fake news opinions as fake claims. We collected both news articles and their applicable claims (short summaries) to account for the variation in text lengths, thereby enhancing the diversity of our dataset.
\\
\begin{figure}[!htb]
\vspace{-0.2in}
\centering
\scalebox{0.45}{
\includegraphics[width = \textwidth]{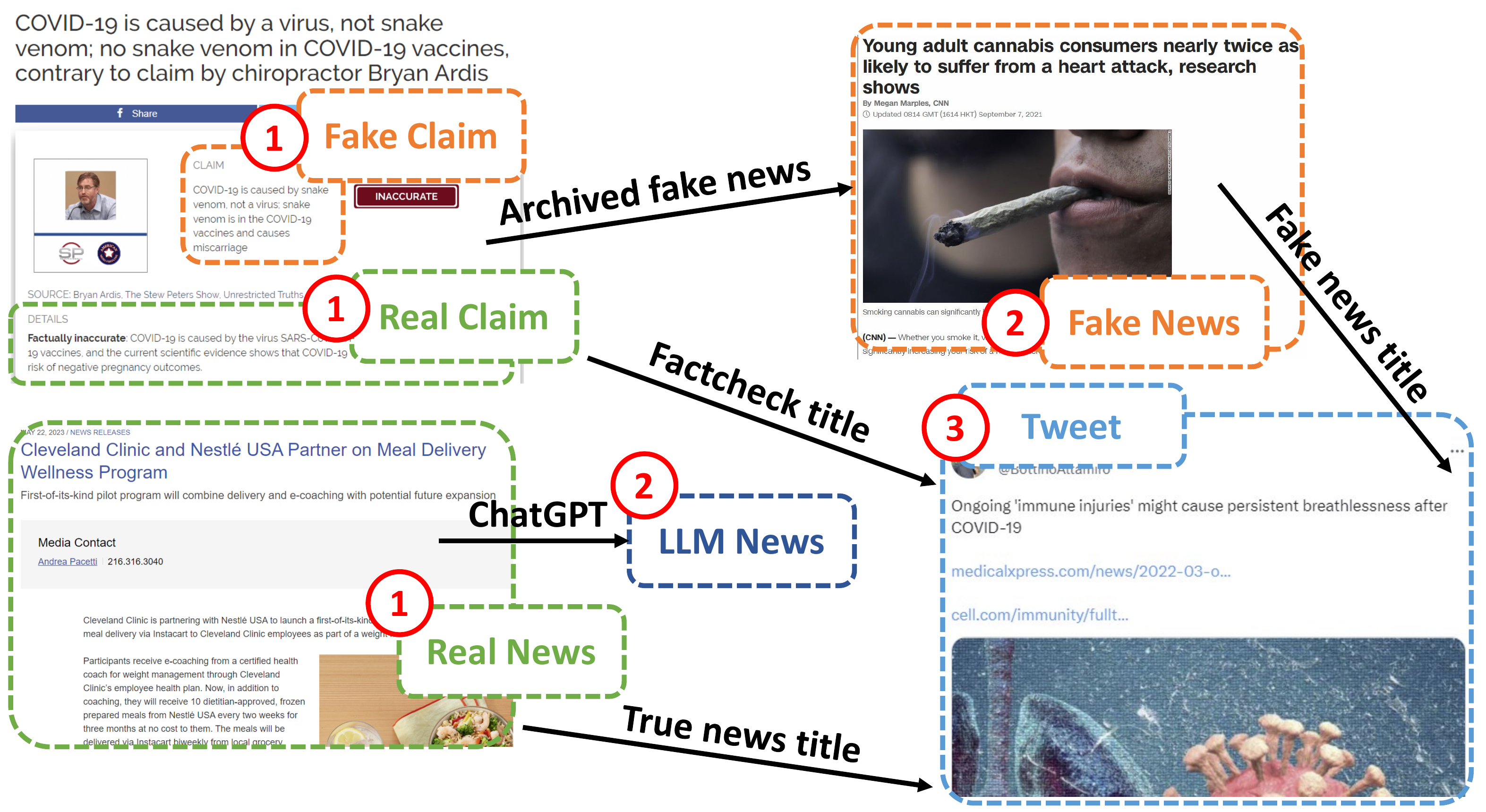}}
\vskip -1em
\caption{The data collection process. The numbers in the red circles indicate the three steps of data collection.}
\vskip -1em
\label{fig:workflow}
\end{figure}
\\
\textbf{Step 1: Content extraction.} In this step, we acquired all the articles from the aforementioned websites, encompassing the text contents, images, and links, spanning from Jan-01-2017 to May-01-2023.
The real news articles can be obtained directly from the news released by the authorities. 
To ensure the dataset's scalability to disease classification tasks, we collected real news containing only one disease label out of a disease list. 
We specifically extracted disease labels that had more than fifty real news articles.
As for claims, in fact-checking articles, fact-checkers typically provide a one-sentence summary of the fake news' opinion, along with their own comments (usually presenting an opposing opinion of the fake news) as illustrated in Fig.~\ref{fig:workflow}. 
In this case, the summaries of news articles identified as "incorrect," "inaccurate," "misinformation," and similar terms by the fact-checkers were labeled as "fake claims," while the corresponding corrections provided by the fact-check articles are considered "real claims."
\\
\textbf{Step 2: Acquisition of human- and LLM-generated fake news.} To assess the effectiveness of fake news detection models against fake content generated by both humans and LLMs, we developed strategies to acquire these two types of fake news.
The fake news articles extracted from Step 1 are human-generated and thus called "human-generated fake news." Additionally, we devised a strategy to simulate adversarial attacks using chatGPT3.5~\cite{chatgpt} on real news articles. Each real news article had a $50\%$ probability of being modified by chatGPT3.5. If chosen for modification, each sentence within the article had a random $10\%$--$50\%$ chance of being altered by providing the prompt "What is the opposite opinion of <the sentence>." These modified sentences were labeled as "fake sentences." Following sentence modification, the attacked article was further refined using the prompt "Refine the language of <the article>." The resulting generated articles were then cleaned up by removing redundant terms such as "the refined version is" or "refinement:" before being labeled as "LLM fake news."
\\
\textbf{Step 3: Real and fake tweet Crawl.}  We crawled tweets spanning from Jan-01-2022 to May-01-2023. This time range was chosen to comply with the size limitation specified in the Tweet Developer Agreement~\cite{tweet2023developer}, as collecting tweets from the past six years would exceed the allowed size. Additionally, this range does not overlap with the time periods covered by the previous datasets in Tab. \ref{tab:compare_dataset}.  Our method of tweet crawling is intrinsically tied to the corresponding news articles that we've crawled. Specifically, we employed the titles of these news articles as key phrases to retrieve related tweets. If the news title is for real news, we categorize the resulting tweets as real. Conversely, if the news title is for fake news, the collected tweets are classified as fake. Owing to the Twitter Developer Agreement~\cite{tweet2023developer}, we might not manipulate tweets by LLMs and could only release the tweet IDs, along with a code that enables users to retrieve the full content of the tweets by these IDs. 

\vspace{-5pt}
\section{Benchmark Tasks \& Statistics}
\begin{figure*}[h]
\centering
\begin{subfigure}{0.32\textwidth}
    \includegraphics[width=\textwidth]{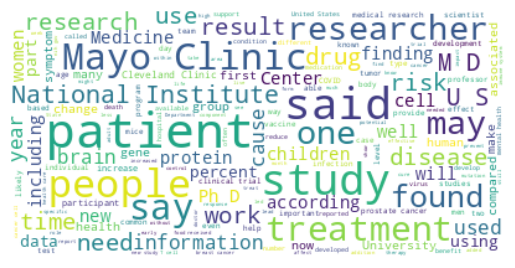}
    \caption{True News.}
    \label{fig:news_fre}
\end{subfigure}
\hfill
\begin{subfigure}{0.32\textwidth}
    \includegraphics[width=\textwidth]{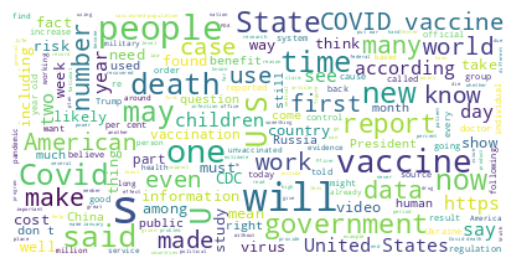}
    \caption{Human-generated Fake News.}
    \label{fig:fakenews_fre}
\end{subfigure}
\hfill
\begin{subfigure}{0.32\textwidth}
    \includegraphics[width=\textwidth]{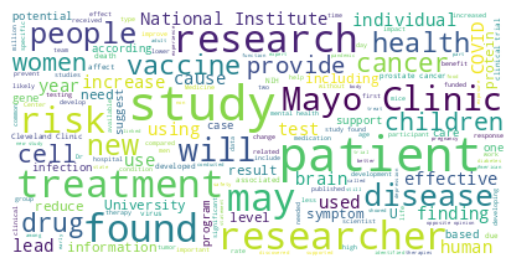}
    \caption{LLM-generated Fake News.}
    \label{fig:LLMfakenews_fre}
\end{subfigure}
\vspace{-0.1in}
\caption{Word Cloud of the date in Med-MMHL.}
\label{fig:wordcloud}
\vspace{-0.15in}
\end{figure*}

\begin{table}[h!]
\small
\centering
\caption{Statistics of benchmark tasks on Med-MMHL, where ``fake news'' is human-generated fake news, ``sent'' is an abbreviation of ``sentence''. 
Since a text might have more than one image, the ``\#Image'' can be larger than ``w/image''.}
\label{tab:stat_info}
\vskip -1em
\begin{tabular}{clccc}
    \toprule
\textbf{Tasks}    &\textbf{Data Type} & \textbf{Count} & \textbf{W/ Image} & \textbf{\# Image}\\
    \midrule
\multirow{5}{*}{\makecell{Fake news \\detection}}&Real news &3,455&/&/\\
&Fake news &469&/&/\\
&LLM fake news &2,095&/&/\\
&Real claim &2,283&/&/\\
&Fake claim &3,567&/&/\\

\midrule
\multirow{2}{*}{\makecell{LLM-generated fake \\sent detection}}&Real sent&41,365&/&/\\
&LLM fake sent&17,608&/&/\\
\midrule
\multirow{4}{*}{\makecell{Multimodal\\ fake news \\detection}}&Real news &4,554&1,338&1,747\\
&Fake news&469&396&5,496\\
&Real claim &643&641&749\\
&Fake claim &1,135&1,102&1,102\\

\midrule
\multirow{2}{*}{\makecell[tl]{Fake tweet \\detection}}&Real tweet&7,738&/&/\\

&Fake tweet &6,927&/&/\\
\midrule

\multirow{2}{*}{\makecell{Multimodal tweet \\detection}}&Real tweet&7,738&1,334&1,385\\
&Fake tweet &6,927&639&763\\
    
\bottomrule
\end{tabular}
\vskip -15pt
\end{table}

We propose and benchmark five different tasks that cover a range of challenges, each involving one or more of four types of inputs: long articles, claims(short articles, as described in Sec.~\ref{sec:data_crawl}), tweets, and multimodal data. The statistics for each task are summarized in Tab. \ref{tab:stat_info} and each task is detailed below.

\textbf{Fake news detection} task specifically concentrates on text-only tasks, encompassing both articles and claims. Images are excluded from this task due to the lack of specific image associations with the text generated by the LLM. Notably, the real news articles used for generating LLM fake news are not included in this task.

\textbf{LLM-generated fake sentence detection} task is designed to evaluate the vulnerability to adversarial attacks introduced by LLM. It goals to assess a model's ability to distinguish between real sentences and LLM-generated fake ones. Therefore, this task excludes human-generated fake sentences.

\textbf{Multimodal fake news detection} aims to investigate ways to enhance the detection of misinformation by leveraging multimodal resources. The specific approach employed for claim filtering is elaborated upon in Appendix~\ref{app:claim_filtering}.

\textbf{Fake tweet detection} and \textbf{multimodal tweet detection} tasks are devised to address the distinctive writing style exhibited in tweets as compared to news articles. In order to fully leverage the available data, all collected tweets are included in both tasks, despite the relatively small number of tweets accompanied by images. 

\textbf{Other applicable tasks on Med-MMHL.} Though we benchmark the above five tasks, Med-MMHL can also be applied to other tasks, given its data diversity. For example, a misinformation detection model can be trained using real news and human-generated fake news, and subsequently employed to identify LLM-generated fake news. Moreover, though our LLM-generated fake sentence detection task excludes the news context, Med-MMHL supports training a model for a more fine-grained misinformation detection at the sentence level.

\section{Misinformation Detection in Medical Domain} 
To demonstrate the main utility of the proposed dataset and evaluate the existing fake news detection methods, we conduct comparative experiments on the misinformation detection task. 
\subsection{Baseline Methods}
We consider seven text-only baseline models and two multimodal baseline models. Specifically, among the seven text-only models, four incorporate language transformer layers pretrained on long articles, two utilize language transformer layers pretrained on sentences, and one is trained using our own dataset. The multimodal models include state-of-the-art pretrained modules for both texts and images. The details of the baselines can be found in Appendix~\ref{app:baseline}. 

\subsection{Implementation Detail}
The dataset is split into training, validation, and testing datasets with a ratio of $7:1:2$. 
Each baseline model comprises a pre-trained module for learning hidden representations and a trainable module for fine-tuning the specific downstream task. During training, the parameters of the pre-trained models remain fixed, and they are utilized to extract hidden representations from the texts and images. A trainable two-layer feedforward neural network module maps the hidden representations to the downstream task. The optimizer is Adam, with a learning rate of $1e^{-5}$ for all the models. The maximum number of epochs is 100 with a 15-step patience. The dropout rate is 0.1. Due to the limitation of our computation resources, the batch size is 4. We adopt commonly used metrics in related areas: Accuracy, Precision, Recall, F1 and Macro F1.

\subsection{Experimental Results} 
We conducted fake news detection and fake-news-related tweet detection experiments on the proposed Med-MMHL dataset. The experiment results are provided in Table~\ref{tab:res_textonly} and Table~\ref{tab:res_tweet}. The metrics used for evaluation include accuracy (Acc), precision (Prc), recall (Rcl), f1-score (F1), and marco f1-score (F1-ma).
We observe that 
\textbf{(i) Pretrained transformer-based methods perform better than simple methods}, as they are more powerful in capturing contextual information better. However, as the dataset is quite imbalanced, the models tend to generate many fake positive cases. Thus, the recall value is lower than the accuracy and precision value. 
\textbf{(ii) FN-BERT performs best} on document-level fake news/tweet detections among all baselines. This indicates the importance of related fake news classification knowledge.
\textbf{(iii)} Although baseline methods show strong performance in detecting fake news, \textbf{the performance of the LLM sentence detection task is unsatisfactory}. It is easier to detect LLM-generated fake news than detect LLM-generated fake sentences, mostly because the generated fake news is entirely opposite in intention to real news, but the generated fake sentences are only partially opposite in intention to real news. Therefore, learning to detect LLM-generated fake sentence detection is an important area for further research.

\begin{table}[h]
\small
\centering
\caption{Baseline methods performance for fake news detection on Med-MMHL.}
\label{tab:res_textonly}
\vskip -1em
  \begin{tabular}{clllll}
    \toprule
    \textbf{Model} & \textbf{Acc} & \textbf{Prc} & \textbf{Rcl} & \textbf{F1}& \textbf{F1-ma}\\
\hline
\multicolumn{6}{c}{\makecell{Fake news detection (both human and LLM-generated fake news)}}\\
\hline
dEFEND&89.174\%&97.361\%&81.240\%&88.573\%&89.144\%\\
BERT&95.657\%&97.702\%&\textbf{93.791\%}&95.707\%&95.657\%\\

BioBERT&94.941\%&98.084\%&91.993\%&94.941\%&94.941\%\\%
Funnel &94.604\%&98.668\%&90.768\%&94.553\%&94.603\%\\

FN-BERT&\textbf{95.784\%}&\textbf{99.472\%}&92.320\%&\textbf{95.763\%}&\textbf{95.784\%}\\

\hline
\multicolumn{6}{c}{\makecell{LLM-generated fake sentence detection}}\\
\hline
dEFEND&92.183\%&88.168\%&85.264\%&86.692\%&90.579\%\\
SentenceBERT&\textbf{96.040\%}&\textbf{96.583\%}&\textbf{89.917\%}&\textbf{93.131\%}&\textbf{95.175\%}\\
DistilBERT&95.149\%&95.050\%&88.355\%&91.581\%&94.087\%\\

\hline
\multicolumn{6}{c}{\makecell{Multimodal fake news detection (only human-generated fake news)}}\\
\hline
CLIP&\textbf{96.324\%}&86.921\%&\textbf{99.377\%}&\textbf{92.732\%}&\textbf{95.136\%}\\
VisualBERT&96.103\%&\textbf{89.881\%}&94.081\%&91.933\%&94.682\%\\
  \bottomrule
\end{tabular}
\vskip -1em
\end{table}

\begin{table}[h]
\small
\centering
\caption{Baseline methods performance for fake-news-related tweet detection on Med-MMHL.}

\label{tab:res_tweet}
\vskip -1em
  \begin{tabular}{clllll}
    \toprule
    \textbf{Model} & \textbf{Acc} & \textbf{Prc} & \textbf{Rcl} & \textbf{F1}& \textbf{F1-ma}\\
\hline
\multicolumn{6}{c}{\makecell{Fake tweets detection (only human-generated fake news)}}\\
\hline
dEFEND&96.897\%&98.868\%&94.517\%&96.643\%&96.880\%\\
BERT&98.056\%&99.552\%&96.318\%&97.908\%&98.046\%\\
BioBERT&97.988\%&\textbf{99.775\%}&95.957\%&97.828\%&97.977\%\\
Funnel&98.158\%&99.701\%&96.390\%&98.018\%&98.149\%\\
FN-BERT&\textbf{98.602\%}&99.339\%&\textbf{97.690\%}&\textbf{98.507\%}&\textbf{98.596\%}\\

\hline
\multicolumn{6}{c}{\makecell{Multimodal fake tweets detection (only human-generated fake news)}}\\
\hline
CLIP&\textbf{97.954\%}&99.256\%&\textbf{96.387\%}&\textbf{97.801\%}&\textbf{97.944\%}\\
VisualBERT&96.404\%&\textbf{99.403\%}&92.620\%&95.985\%&96.364\%\\
  \bottomrule
\end{tabular}
\vskip -1em
\end{table}

\section{Conclusion}

Medical misinformation significantly affects individuals and societies, necessitating effective detection methods. However, existing datasets have limitations: overlooking visual information, focusing solely on COVID-19, or ignoring LLM-generated misinformation. To address these limitations, we introduce Med-MMHL, a multi-modal dataset for detecting misinformation in the broader medical field, incorporating both human and LLM-generated fake data across multiple diseases. Additionally, Med-MMHL extends its diversity by incorporating data from news and tweets. We also comprehensively analyze the dataset's characteristics at text and sentence levels. Finally, we establish a benchmark for misinformation detection at sentence, document, and multi-modal levels, laying the groundwork for future research in this critical domain.

\bibliographystyle{ACM-Reference-Format}
\bibliography{sample-base,sigir}

\clearpage
\appendix

\section{Appendix}

\subsection{Data Overview}


\begin{table*}[h]
\small
\centering
\caption{Statistics between diseases and news/tweets.}
\vskip -1em
  \begin{tabular}{clllll lllll lllll l}
    \toprule
    \textbf{Information Type} & 
    \mcrot{1}{l}{30}{\textbf{anemia}} & 
    \mcrot{1}{l}{30}{\textbf{arthritis}} & 
    \mcrot{1}{l}{30}{\textbf{asthma}} & 
    \mcrot{1}{l}{30}{\textbf{cancer}} & 
    \mcrot{1}{l}{30}{\textbf{covid}} & 
   \mcrot{1}{l}{30}{\textbf{diabetes}} & 
   \mcrot{1}{l}{30}{\textbf{epilepsy}} & 
   \mcrot{1}{l}{30}{\textbf{flu}} & 
   \mcrot{1}{l}{30}{\textbf{headache}} & 
   \mcrot{1}{l}{30}{\textbf{hypertension}} & \mcrot{1}{l}{30}{\textbf{inflammation}} & \mcrot{1}{l}{30}{\textbf{monkeypox}} & 
   \mcrot{1}{l}{30}{\textbf{parkinson}} & 
   \mcrot{1}{l}{30}{\textbf{pneumonia}} & 
   \mcrot{1}{l}{30}{\textbf{stroke}} & 
   \mcrot{1}{l}{0}{\textbf{Total}}\\
    \midrule
Real news&64&85&148&1,410&859&332&48&740&70&55&282&44&81&50&286&4,554\\
Fake news&0&1&0&27&304&1&2&114&1&0&4&3&0&2&10&469\\
LLM fake news&18&35&62&615&462&161&23&339&30&25&135&19&39&11&121&2,095\\
True claims&3&4&7&190&1,619&31&2&362&11&1&12&7&4&9&21&2,283\\
False claims&5&6&10&269&2,557&38&3&575&14&2&14&19&6&15&34&3,567\\
\midrule
\textbf{Total news}&91&133&227&2,560&5,836&569&83&2,152&127&84&452&94&132&88&481&12,968\\
\midrule
Real tweets&53&15&28&540&1,161&152&29&2,095&36&21&174&8&35&17&106&7,738\\
Fake tweets&0&0&0&120&2,547&0&1&2,436&0&0&2&1,799&0&0&21&6,927\\
\midrule
\textbf{Total tweets}&53&15&28&660&3,708&152&30&4,531&36&21&176&1,807&35&17&127&27,633\\
  \bottomrule
\end{tabular}
\label{tab:stat_disease}
\vskip -1em
\end{table*}

\subsubsection{Relations to Multiple Disease}
Although we did not provide specific disease labels for the articles/tweets, we conducted a statistical analysis based on diseases. As indicated in Table \ref{tab:stat_disease}, we examined fifteen disease categories that contained more than fifty real news articles. The statistical findings reveal that the number of real news articles is relatively evenly distributed across various types of diseases. In contrast, fake news articles and tweets tend to concentrate on "hotspot" topics such as Covid-19 and Monkeypox.

\subsection{Data Analysis}
We analyze our dataset in two-fold: text-level and embedding-level. We detail these two folds below.

\begin{figure*}[h]
\centering
\begin{subfigure}{0.47\textwidth}
    \includegraphics[width=\textwidth]{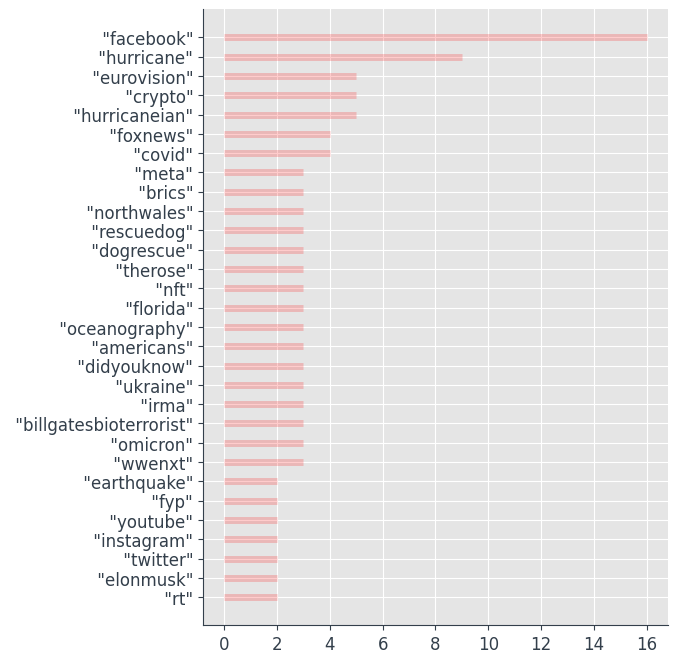}
    \caption{Fake News.}
    \label{fig:fakehashtag}
\end{subfigure}
\hfill
\begin{subfigure}{0.47\textwidth}
    \includegraphics[width=\textwidth]{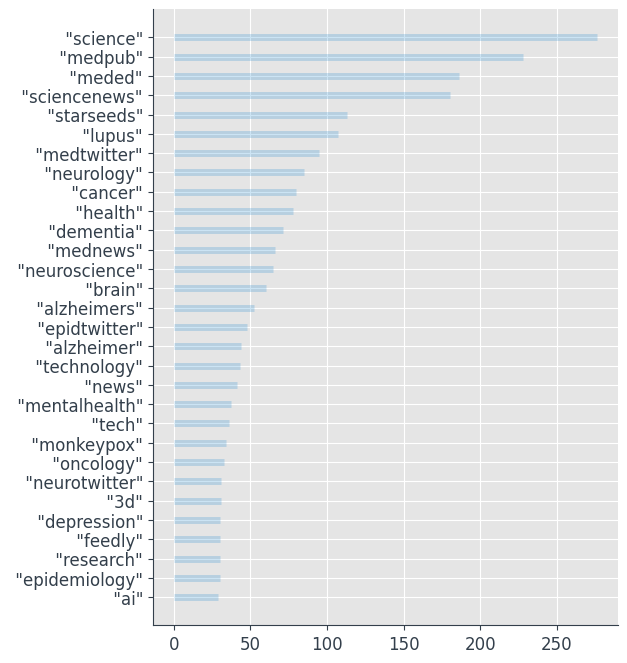}
    \caption{True News.}
    \label{fig:truehashtag}
\end{subfigure}
\vskip -1em
\caption{Frequency of hashtags in tweets about fake and true news articles.}
\vskip -1em
\label{fig:figures}
\end{figure*}

\noindent\textbf{text-level.} To understand the topic difference between the tweets of fake and real news, we analyze the top 30 frequent hashtags in tweets related to fake and true news articles, respectively. The frequency of hashtags in tweets related to fake and real news articles is shown in Figure~\ref{fig:fakehashtag} and Figure~\ref{fig:truehashtag}, respectively. We ﬁnd that the hashtag distributions of tweets about fake and real news articles are quite different. While the hashtags in tweets about true news articles are mainly related to healthcare, those in tweets about fake news cover more diverse topics, including social media (\#facebook, \#foxnews) and natural disasters (\#hurricane, \#earthquake).

\begin{figure}[!htb]
\centering
\scalebox{0.45}{
\includegraphics[width = \textwidth]{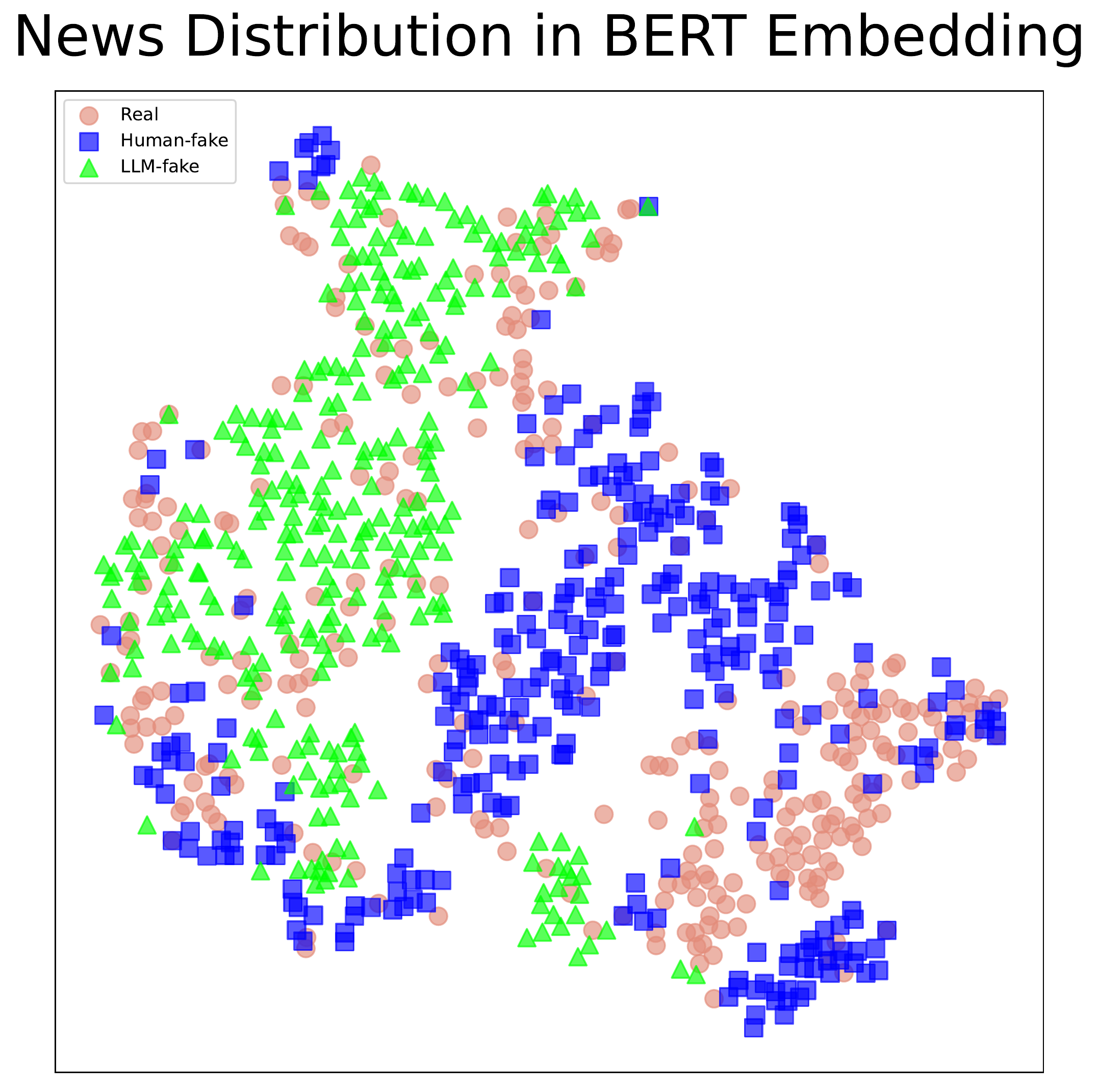}}
\vskip -1em
\caption{A t-SNE figure of randomly sampled 300 real news, 300 human-fake news, and 300 LLM-fake news.}
\vskip -1em
\label{fig:news_emb}
\end{figure}

\noindent\textbf{embedding-level.} In terms of news, we categorized our crawled content into three distinct sources: real, human-generated fake, and Language Learning Model (LLM)-generated fake news. As depicted in Figure~\ref{fig:news_emb}, we randomly selected 300 news articles from each of these categories and analyzed them using BERT embeddings. However, our analysis reveals that the BERT embeddings struggle to distinguish between real, human-fake, and LLM-fake news due to significant overlaps in these categories.

This observation highlights the significance of researching methodologies to accurately discern these three distinct sources of news. Moreover, our analysis shows minimal overlap between LLM-fake news and human-fake news, suggesting that a model adept at identifying human-fake news might not necessarily be effective at detecting LLM-generated misinformation, and vice versa. This calls for an approach that can adapt to these distinct categories effectively. 

\begin{figure}[!htb]
\centering
\scalebox{0.45}{
\includegraphics[width = \textwidth]{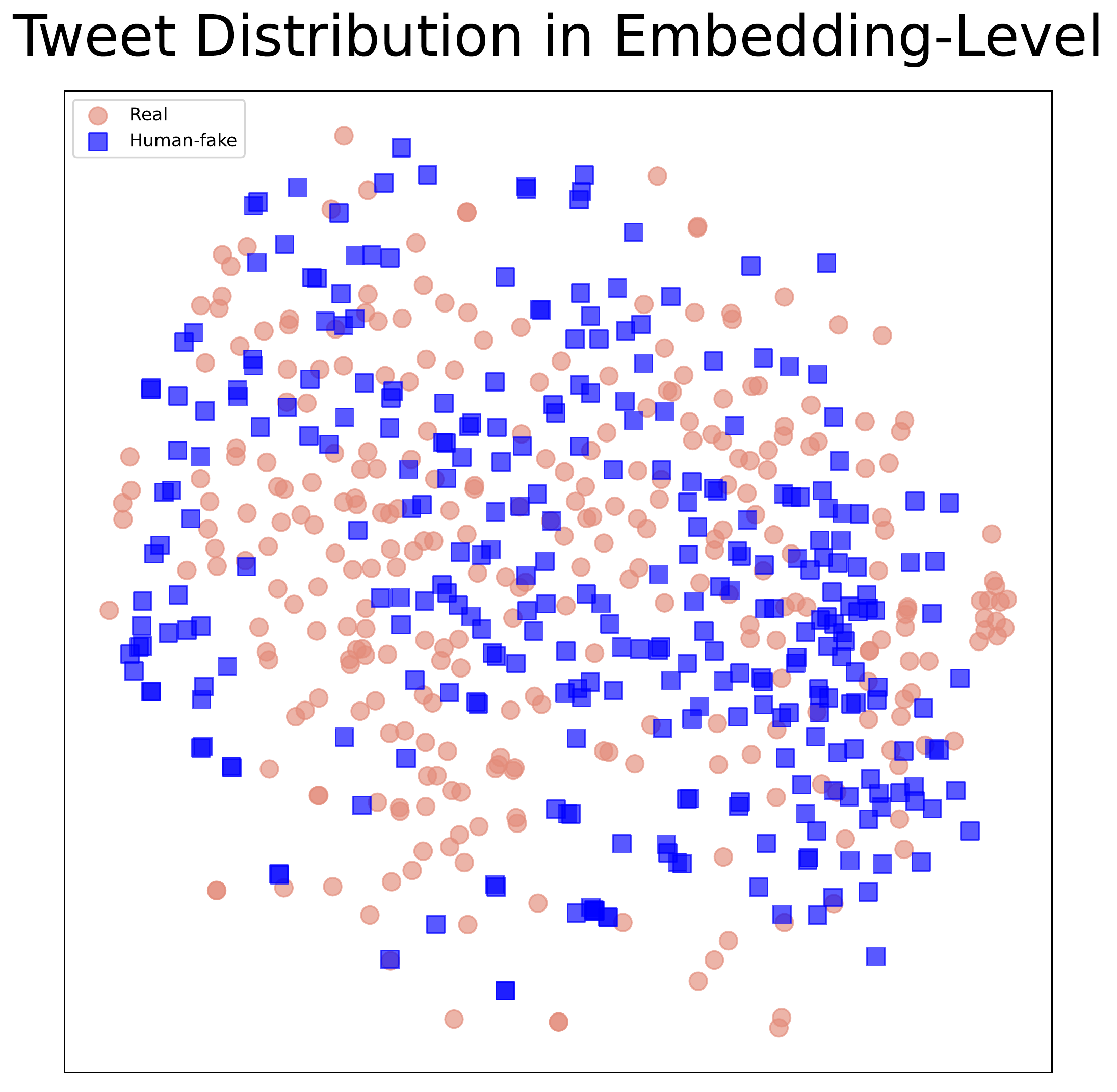}}
\vskip -1em
\caption{A t-SNE figure of randomly sampled 300 real tweets, 300 human-fake tweets. Due to the Tweet Develop Policy~\cite{tweet2023developer}, we cannot use ChatGPT to generate LLM-fake tweets.}
\vskip -1em
\label{fig:tweet_emb}
\end{figure}

Correspondingly, we categorized the crawled tweets into two primary sources: real tweets and human-fake tweets. Due to the constraints imposed by Twitter's Developer Policy~\cite{tweet2023developer}, the generation of LLM-fake tweets is not permissible. As a result, we randomly sampled 300 tweets from both sources for our analysis, as illustrated in Figure~\ref{fig:tweet_emb}. For analysis, we utilized TweetBERT embeddings~\cite{qudar2020tweetbert}. However, the figure shows that TweetBERT embeddings struggle to clearly demarcate between real and human-fake tweets, demonstrating significant overlap. This underlines the importance of exploring further research methodologies to distinguish these two categories accurately. 

\subsection{Multimodal Claim Filtering} \label{app:claim_filtering}

By looking at the images of the real news, fact-check, and fake news articles, we notice a pattern where real news articles often incorporate decorative images sourced from the internet, while fake news articles frequently utilize screenshots of social media or videos. This stark contrast makes it relatively straightforward to distinguish between true and fake news. 
However, in the case of fact-check articles, we observe that "AFPFactCheck" tends to use screenshots, while "CheckYourFact" and "PolitiFact" lean towards using decorative images. Consequently, we included the true claims from "AFPFactCheck" and the false claims from "CheckYourFact" and "PolitiFact" as part of the multimodal fake news detection task. 
This ensures that the models trained on our dataset do not get misled by features that are irrelevant to the content of the articles.

\subsection{Baseline Models}\label{app:baseline}

The following baseline fake news detection methods are considered for medical misinformation detection:
\begin{itemize}[leftmargin=*]
    \item BERT~\cite{bert}: A bi-directional transformer model pretrained on a large corpus of English data in a self-supervised fashion.
    \item BioBERT~\cite{biobert}: A sentence-transformers model built with medical dataset for fact-checking of online health information.
    \item Funnel Transformer~\cite{dai2020funneltransformer}: An efficient bidirectional transformer model by applying a pooling operation after each layer, akin to convolutional neural networks, to reduce the length of the input.
    \item FN-BERT~\cite{fnbert}: A BERT-based model recently finetuned on a Fake news classification dataset in 2023. 
    \item sentenceBERT~\cite{sentencebert}: A sentence representation learning model pretrained using Siamese and triplet network structures.
    \item distilBERT~\cite{distilbert}: A dual-encoder then dot-product scoring architecture BERT model. The version employed in this paper is pre-trained with the TAS-Balanced method on the MSMARCO standard.
    \item dEFEND~\cite{shu2019defend} utilizes the hierarchical attention network to model article content for misinformation detection.
    \item CLIP~\cite{clip}: A multi-modal vision and language model pretrained on 400 million image-text pairs.
    \item VisualBERT~\cite{visualbert}: A multi-modal vision and language model. It uses a BERT-like transformer to prepare embeddings for image-text pairs. 
\end{itemize}

\end{document}